\documentstyle[aps,prl,floats,graphicx]{revtex}
\begin{document} 
\draft 
\twocolumn[\hsize\textwidth\columnwidth\hsize\csname 
@twocolumnfalse\endcsname

%\preprint{Preprint } 
\title{ Domain structure of superconducting ferromagnets}  
\author{E.B. Sonin}

\address{Racah Institute of Physics, Hebrew University of
Jerusalem, Jerusalem 91904, Israel } 

\date{\today} \maketitle

\begin{abstract}
In superconducting ferromagnets the equilibrium domain structure is
absent in the Meissner state, but appears in the spontaneous vortex
phase (the mixed state in zero external magnetic field), though with a
period, which can essentially exceed that in normal ferromagnets.
Metastable domain walls are possible even in the Meissner state. The
domain walls create magnetostatic fields near the sample  surface,
which can be used for experimental detection of domain walls.

\end{abstract}

%\eject
] 
%\narrowtext
%\twocolumn

Recently there has been a growing interest to materials, in which
superconductivity and ferromagnetism coexist
\cite{felner,Nat1,Nat2,Chu,Tallon}. A number of  unusual phenomena and
structures have been predicted and observed, spontaneous vortex phase as an
example\cite{GBV,SF}. But the theory mostly addressed macroscopically uniform
structures, whereas ferromagnetic materials, even ideally uniform, inevitably
have a domain structure, which is a ground-state property of ferromagnets. So
a further progress in studying materials with coexisting ferromagnetism and
superconductivity requires an analysis of the domain structure. The present
work is the first step in this direction.

An object of the study is a material, in which the
magnetic transition occurs earlier, i.e. at a higher temperature, than
the superconductivity onset. This was called ``superconducting
ferromagnet'' \cite{SF}, in contrast to ``ferromagnetic (or magnetic)
superconductors'' where the superconductivity sets in {\em before} the
magnetic transition, which have been studied mostly in the past
\cite{BulBuz}.  Competition of ferromagnetism and superconductivity may
result in various structures with the magnetic moment rotating
in space  (spiral structures, cryptoferromagnetism and so on). This
also can be considered as a ``domain structure'', but with a
period determined by intrinsic properties of materials. However, our
goal is the domain structure due to magnetostatic fields
generated by nonzero average bulk magnetization $\vec M$. In this
case the domain size depends on a sample size. We shall consider
type-II superconductivity, bearing in mind ruthenocuprates
\cite{felner,Chu,Tallon}, which are type II high-$T_c$ superconductors.

Before analyzing the domain structure it is useful to summarize the
magnetic properties of a single-domain superconducting ferromagnet. The
total free energy of the superconducting ferromagnet can be written as
\cite{SF}
\begin{equation}
F(\vec M, \vec B)=f_E+K +{
(\vec B- 4\pi \vec M)^2
\over 8\pi } +{2\pi \lambda^2 \over c^2}j_s^2 ~,
    \label{F-tot}\end{equation}
where   $\lambda$ is the London penetration depth and $\vec B$ is the
magnetic induction. The energy $f_E(M,\nabla \vec M)$ is the exchange
energy, which depends on the absolute value of $M$ and on gradients of
$\vec M$. As a rule \cite{LL}, in magnetic
materials this is the largest energy, which fixes $M$. The anisotropy 
energy $K(\vec M/M)$ is smaller and depends on the direction of $\vec M$.
We shall consider a stripe magnetic structure, which is possible only if
$K$  essentially exceeds  the magnetostatic energy $\sim M^2$ \cite{LL}.
The latter is determined by the magnetic field
$\vec H=\vec B- 4\pi \vec M$ [the third term in Eq. (\ref{F-tot})]. The
expression Eq. (\ref{F-tot}) includes also the kinetic energy related to
the superconducting current
\begin{equation}
\vec j_s={c\Phi_0 \over 8\pi^2 \lambda^2}\left(\vec \nabla \varphi -
{2\pi \vec A \over \Phi_0}\right) ~,
    \label{j-s}\end{equation}
where  $\Phi_0$ is the magnetic-flux quantum,  $\varphi$ is the phase
of the superconducting order parameter, and the vector potential
$\vec A$ determines the magnetic induction $\vec B =\vec \nabla
\times \vec A$. The kinetic
energy of superconducting currents is absent in a normal
ferromagnet.

Minimization of the energy with respect to the
vector potential
$\vec A$ yields the Maxwell equation
\begin{equation}
{4\pi \over c}\vec j_s
 =\vec \nabla \times (\vec B -4\pi \vec M)
=\vec \nabla \times \vec B ~.
    \label{Max} \end{equation}
Together  with the equation 
\begin{equation}
\vec \nabla \times \vec j_s=- \frac{c}{4\pi \lambda^2} \vec B
     \label{rot-j}\end{equation}
this  yields the London equation which determines $\vec B$: 
\begin{equation}
\lambda^2 \vec \nabla \times [\vec \nabla \times \vec B] 
+ \vec B =0~.
    \label{Lon-mu} \end{equation}
Here we took into account that  $\vec \nabla \times \vec M=0$ inside
domains. In contrast to Ref. \cite{SF}, we neglect
the differential  susceptibility ($M$ does not depend on a magnetic
field), which renormalizes the London penetration depth.

\begin{figure}[!b]
  \begin{center}
    \leavevmode
    \includegraphics[width=0.9\linewidth]{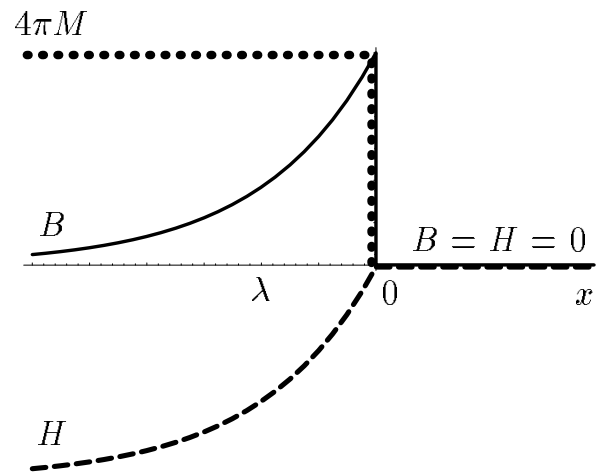}
    \bigskip
    \caption{Magnetic induction $B$ (solid line), magnetic field $H$ 
    (dashed line), and $4\pi M$ (dotted line) at the boundary between
a superconducting ferromagnet ($x<0$) and vacuum ($x>0$). }
  \label{fig1}
  \end{center}
  \end{figure}

These equations and the boundary conditions at the sample boundary
(continuity of the tangential component of $\vec H$ and of the normal
component of $\vec B$) yield the distribution of $\vec B $ and
$\vec H$. This distribution is shown in Fig. \ref{fig1}
for the case of $\vec M$ parallel to the sample boundary and for zero
external magnetic field. The magnetic induction and the related magnetic
flux exist only in the layer of the thickness $\lambda$. Meissner
currents in this layer screen the internal field $4\pi \vec M$, as well
as they screen the external magnetic field in a nonmagnetic
superconductor.

\begin{figure}%[!h]
  \begin{center}
    \leavevmode
    \includegraphics[width=0.9\linewidth]{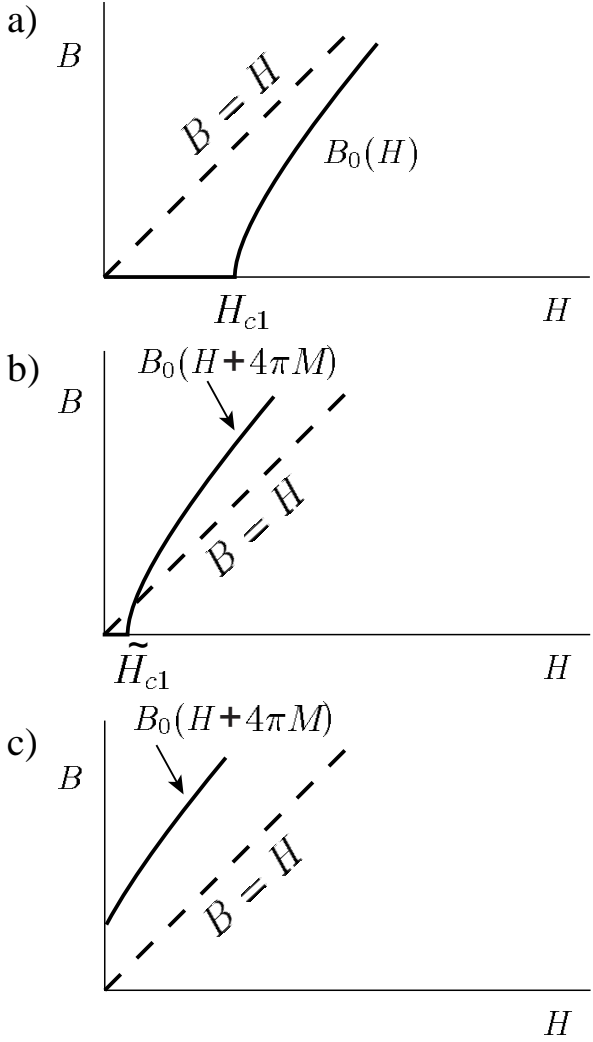}
    \bigskip
    \caption{Magnetization curve: a) nonmagnetic type-II
superconductor; b) superconducting ferromagnet, $4\pi M < H_{c1}$;
c)  superconducting ferromagnet, $4\pi M > H_{c1}$. }
  \label{fig2-1}
  \end{center}
  \end{figure}

Let us consider now the mixed state of the superconducting
ferromagnetic, in which vortices (magnetic fluxons) are present in
the bulk. Since ferromagnetism does not affect the London equation
(\ref{Lon-mu}), one expect the same magnetic-induction distribution in
the mixed state as for nonmagnetic type II superconductors  \cite{SF},
and the free energy is given by
\begin{equation}
F_m(\vec M, \vec B)=f_E+K
+2\pi M^2 -\vec B\cdot \vec M +F_0(B) ~,
    \label{F-mix}\end{equation}
where $F_0(B)$ is the free energy of a nonmagnetic type II
superconductor, and $\vec B$ now is the magnetic
induction averaged over the vortex-array cell. The energy $F_0(B)$
contains both the magnetic energy $B^2/8\pi$ and the kinetic energy of the
superconducting currents inside the vortex cell. Determining the
magnetic field 
\begin{equation}
\vec H = 4\pi {\partial F_m \over \partial \vec B}=
 4\pi {\partial F_0 \over \partial \vec B}-4\pi \vec M~,
   \label{h-b} \end{equation}
we see that the magnetization curve of a superconducting ferromagnet
is described by
$B=B_0(|\vec H +4\pi
\vec M|)$  where $B_0(H)$ is the equilibrium magnetization curve for
a nonmagnetic type II superconductor \cite{SF} (Fig. \ref{fig2-1}a).
Note that in this relation the magnetic field $\vec H $ has a
different physical meaning from that used in the
Meissner state. For the Meissner state we introduced
$\vec H =\vec B -4\pi \vec M$, where the moment $\vec M$ originates
from ``molecular'' currents responsible for ferromagnetism, the
superconducting currents being treated as external currents. In the
mixed state, which is considered now, it is more convenient to
define the magnetic field as
$\vec H =\vec B -4\pi (\vec M +
\vec M_s)$, i.e. the definition includes also the diamagnetic moment
$\vec M_s=(\vec B_0-\vec H)/4\pi$ of the superconducting currents
circulating around vortex lines in the mixed state. Thus these
currents are treated in the same manner as molecular currents
responsible for ferromagnetism. 

Figure \ref{fig2-1}b shows that in a superconducting ferromagnet the
Meissner state ($B=0$) exists until
$H+4\pi M < H_{c1}$, where $H_{c1}= (\Phi_0 / \lambda ^2)\ln
(\lambda/\xi)$ is the lower critical field in a nonmagnetic superconductor and $\xi$ is the
coherence length, which determines the vortex core size. So ferromagnetism
decreases the lower critical field $\tilde H_{c1}= H_{c1}-4\pi M$. If
$4\pi M > H_{c1}$, the Meissner state is absent (Fig.
\ref{fig2-1}c) and the superconducting ferromagnet is in the mixed state
with vortices penetrating into it even in zero external field $H=0$. This
is  {\em spontaneous vortex phase} with nonzero magnetic induction $B=
B_0(4\pi M)$ in the bulk.

\begin{figure}%[!h]
  \begin{center}
    \leavevmode
    \includegraphics[width=0.9\linewidth]{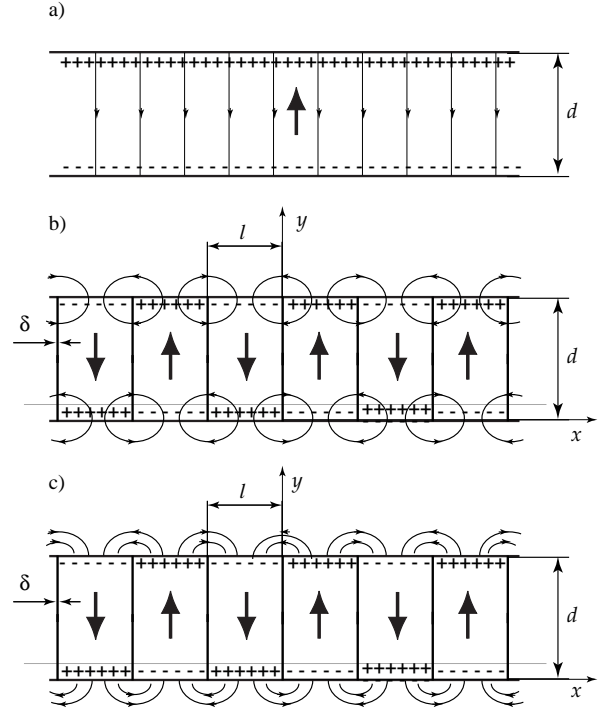}
    \bigskip
    \caption{Domain structure in normal and superconducting
ferromagnets. The thick arrows show directions of the magnetic
moment $\vec M$, the thin lines with arrows are force lines of the
magnetostatic field $\vec H$. The magnetic charges are shown by + and
-.   a) A single-domain structure. In the whole bulk $B=0$ and
$\vec H=-4\pi \vec M$. b) A stripe domain structure in a normal
ferromagnet. The magnetostatic fields are present in areas
$\sim l^2$ inside and outside the sample. In the rest parts of
domains $H=0$ and $\vec B=4\pi \vec M$. c) A superconducting ferromagnet
in the spontaneous vortex phase with a rigid vortex array. The
magnetostatic fields appear only in areas
$\sim l^2$ outside the sample. In the bulk of domains $H=0$
and
$B=B_0(4\pi M)$. }
  \label{fig2}
  \end{center}
  \end{figure}

Now let us consider formation of the domain structure in the standard
geometry \cite{LL}: a slab of the  thickness $d$ along the  anisotropy
easy axis $y$ and infinite in directions of the axes $x$  and $z$ (Fig.
\ref{fig2}). We start from a normal ferromagnet. In the absence of an
external magnetic field the average magnetic induction inside the slab
must vanish. Therefore, $B=0$ in a single-domain structure  (Fig.
\ref{fig2}a), and there exists an uniform magnetostatic field $\vec H
=-4\pi \vec M$ in the entire sample, an analog of the electrostatic field
in a charged plane capacitor. This results in a high magnetostatic energy
$H^2/8\pi
\sim M^2$.  However, the domain structure with period $l$ ($l \ll d$)
suppresses this energy in the domain bulk: $\vec H \approx 0$ and
$\vec B =-4\pi \vec M$, except for the area $\sim l^2$ near the sample
boundary (Fig.
\ref{fig2}b). But the average induction still vanishes, since $\vec M$
changes its sign from a domain to a domain. For the stripe
structure one can solve the equations of magnetostatics, $\vec \nabla
\times \vec H =0$ and $\vec \nabla \cdot \vec H =4\pi \rho_M$, exactly
\cite{LL,S}. Here $\rho_M =-\vec \nabla \cdot \vec M$ is the magnetic
charge. The magnetostatic energy per unit volume
of the slab is
\begin{equation}
E_s =0.852 M^2 l^2\times {1\over ld} = 0.852 M^2 {l\over d} ~.
    \end{equation}
This energy is by a factor $l/d$ less  than the magnetostatic energy in
a single-domain structure. However, the domain walls increase the energy. 
The energy
of one domain wall (per unit length along the slab) is $\alpha  K\delta
d$, where $\delta$ is the wall thickness and the  numerical factor
$\alpha$ depends on the detailed definition of $K$
and $\delta$. Its  specification is not essential for the
present  analysis. The domain-wall energy per unit volume of the sample is
\begin{equation}
E_w =\alpha Kd\delta  \times {1\over ld}= \alpha K{\delta \over l}~.
    \end{equation}
The equilibrium value of the period $l$ is determined 
by minimization of the energy $E_w +E_s$ \cite{LL}:
\begin{equation}
l = \sqrt{{\alpha K\over 0.852 M^2}\delta d}~.
   \label{norm-l} \end{equation}

Let us return back to a superconducting ferromagnet. In the Meissner state
the magnetic induction must vanish in the bulk, which is compatible
only with the single-domain structure. Thus {\em the equilibrium domain
structure is impossible in the Meissner state}. However, domains  with 
the changing direction of $\vec M$   can
appear in the spontaneous vortex phase with nonzero $B= B_0(4\pi M)$. 
Like in a normal ferromagnet, the magnetic flux 
$\propto B$ in domains should produce the magnetostatic fields in the
area $\sim l^2$, but in a
superconducting ferromagnet these fields are by the factor
$B_0(4\pi M)/4\pi M$ smaller. We can take it into account introducing
 the effective magnetization $\tilde M = B_0(4\pi M)/4\pi$. Then
the period of the domain structure is given by Eq.
(\ref{norm-l}), where $M$ must be replaced with $\tilde M$.
In the limit of  large $4\pi M \gg H_{c1}$, one has $\tilde M \rightarrow
M$ and the effect of superconductivity on the domain structure vanishes.
In the opposite limit of small $M$, when $4\pi M \rightarrow H_{c1}$, 
$\tilde M$ vanishes and the period $l$ becomes infinite, as it
should be in the Meissner state $4\pi M < H_{c1}$.  
However, this calculation of $l$ assumes that the penetration of the
magnetostatic field into a superconducting ferromagnet is similar to the
penetration into a normal ferromagnet. The assumption is correct if 
rigidity of the vortex array is negligible and the effective penetration
depth is infinite.  We can also consider the opposite limit of a very
rigid vortex array, when the magnetostatic fields penetrate only into the
layer of the thickness $\lambda$. If $\lambda \ll l$, the
penetration of the magnetic flux into a superconductor becomes
insignificant. This increases the magnetic fields outside the sample,
as well as the total magnetostatic energy, by a factor of 2, while
the correspondingly period $l$  decreases by a factor of
$\sqrt{2}$ (Fig. \ref{fig2}c), in analogy with the effect of a
superconducting substrate on a domain size in a ferromagnetic slab
\cite{S}. Thus avoiding a detailed analysis of the vortex and field
pattern in the domains close to the sample border we lose only a
numerical factor of not more than
$\sqrt{2}$. In any case, superconductivity, which coexists with
ferromagnetism in the same bulk, always increases the domain size, in
contrast to the superconductor-ferromagnet bilayer, where
superconductivity shrinks ferromagnetic domains \cite{S}.

\begin{figure}%[!h]
  \begin{center}
    \leavevmode
    \includegraphics[width=0.9\linewidth]{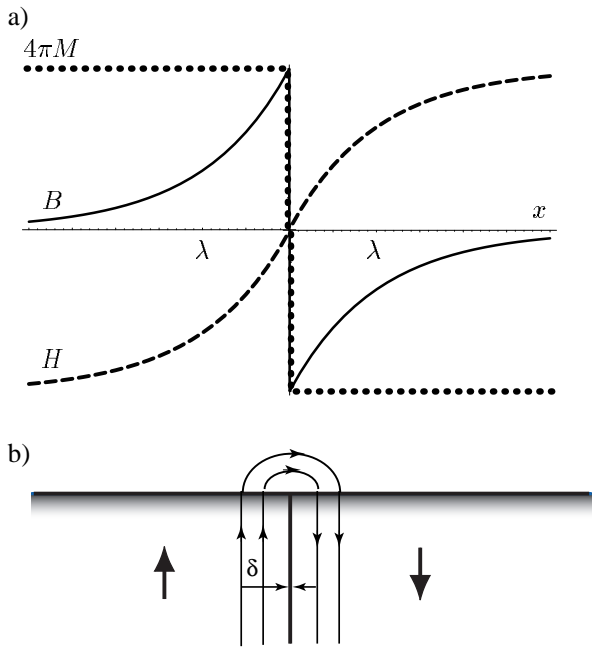}
    \bigskip
    \caption{Domain wall in the Meissner state: a) Magnetic induction $B$ (solid line),
magnetic field $H$ 
    (dashed line), and $4\pi M$ (dotted line) near the domain wall in the 
    superconducting ferromagnet. b) Magnetic flux lines around the exit of the domain wall
(of thickness $\delta$) to the sample surface. }
  \label{fig3}
  \end{center}
  \end{figure}

The absence of the equilibrium domain structure in the Meissner state
does not rule out a possibility of {\em metastable} domain walls, as
topologically stable planar defects.  Domains can appear also
because of disorder, or grain structure. The structure of the domain
wall should be found by solution of the coupled equations of
magnetostatics and the London electrodynamics. We restrict ourselves to
the simplest case, when the London penetration depth $\lambda$
essentially exceeds the domain wall thickness $\delta$. This means that
at the spatial scales of order $\delta$ the domain-wall structure is
governed by large energies [the exchange energy and the anisotropy
energy, see Eq.~(\ref{F-tot})] and is not affected by the magnetostatic
and kinetic energy. On the other hand, at scales $\sim \lambda$ one can
find the distribution of $\vec B$ and $\vec H$ from the London equation
at constant $\vec M$. This is shown for the Bloch domain wall
(the magnetization $\vec M$ rotates in the plane of the wall and does not
produce the magnetostatic charges) in Fig. \ref{fig3}a. Though
our picture corresponds to a 180$^\circ$ wall, a similar
picture is expected for any domain wall. The jump of the tangential
component of the moment $\vec M$ at the wall defines the current sheet,
responsible for a jump of the magnetic induction parallel to the wall,
whereas a possible jump of the normal component of $\vec M$ (a
``charged'' domain wall) would produce a jump of the normal component of
the field $\vec H$.

The magnetic flux on the opposite sides from the domain
wall creates the magnetostatic fields outside the sample, where the
wall meets the sample boundary (Fig. \ref{fig3}b). The magnetic fluxes, which exit from
the sample at two sides from the wall, are equal in magnitude
($4\pi M \lambda $ per unit length along the wall) but opposite in
direction. The magnetostatic field from
domain walls could be used for their experimental detection. At
distances $r \gg \lambda$ from a line, where the wall exits to the
sample boundary, this field is a dipole field  of
the order of
$M\lambda^2 /r^2$.

In summary, the letter presents the first analysis of the domain
structure in superconducting ferromagnets. There is no equilibrium
domain structure in a superconducting ferromagnet in the Meissner
state. In the spontaneous vortex phase the period of the domain structure
may essentially exceed that in the normal ferromagnet. But metastable
domain walls can exist even in the Meissner state.   They generate the
magnetic flux in layers of a thickness
$\lambda$, which can be revealed by magneto-optical methodic. 

The author thanks Yu. Barash, I. Felner, and N. Kopnin for useful
discussions. The work has been supported by the grant of the Israel
Academy of Sciences and Humanities.

\end{document}